\documentstyle[aps,epsf,,multicol,prb]{revtex} 
\renewcommand{\narrowtext}{\noindent\begin{multicols}{2}\noindent
\global\columnwidth20.5pc}
\renewcommand{\widetext}{\end{multicols}
\global\columnwidth42.5pc}  
\renewcommand{\top}[1]{%
 \vskip #1%
 \begin{picture}(290,80)(80,500)%
 \thinlines%
 \put(65,500){\line( 1, 0){255}}\put(320,500){\line( 0, 1){5}}%
 \end{picture}%
}
\newcommand{\bottom}[1]{%
 \vskip #1%
 \begin{picture}(290,80)(80,500)%
 \thinlines%
 \put(330,500){\line( 1, 0){255}}\put(330,500){\line( 0, -1){5}}%
 \end{picture}%
}
\newcommand{\V}[1]{\bf #1}                  
\input{psfig.sty}
\begin{document}
\draft 
{
\title{Auxiliary particle theory of threshold singularities in 
  photoemission and X-ray absorption spectra: test of a conserving
  $T$-\/matrix approximation}

\author{Thomas Schauerte, Johann Kroha and Peter W{\"o}lfle}
\address{Institut f\"ur Theorie der Kondensierten Materie, Universit\"at
  Karlsruhe, Postfach 6980, 76128 Karlsruhe, Germany}
\date{\small November 30, 1999}
\maketitle
\begin{abstract}
  We calculate the exponents of the threshold singularities in the
  photoemission spectrum of a deep core hole and its X-ray absorption
  spectrum in the framework of a systematic many-body theory of slave
  bosons and pseudofermions (for the empty and occupied core level).
  In this representation, photoemission and X-ray absorption can be
  understood on the same footing; no distinction between orthogonality
  catastrophe and excitonic effects is necessary.  We apply the
  conserving slave particle $T$\/-matrix approximation (CTMA),
  recently developed to describe both Fermi and non-Fermi liquid
  behavior systems with strong local correlations, to the X-ray
  problem as a test case.  The numerical results for both
  photoemission and X-ray absorption are found to be in agreement with
  the exact infrared powerlaw behavior in the weak as well as in the
  strong coupling regions.  We point out a close relation of the CTMA
  with the parquet equation approach of Nozi{\`e}res {\em et al.}
\end{abstract}
} 

\narrowtext

\section{Introduction}
\label{introduction}

The core level spectral function $A_d(\epsilon)$ of a localized core 
orbital immersed in a conduction electron sea, as observed in the
photoemission of electrons after X-ray absorption has long been known
to show nonanalytic threshold behavior characterized by fractional
power laws $A_d(\epsilon) \propto \epsilon^{-\alpha_d}$ in the frequency
distance to the threshold $\epsilon = \omega - E_0$. As shown by
Anderson \cite{Anderson67}, this can be understood by considering that
the sudden creation of a deep hole in the electronic core of an ion in
a metal (or the filling of an empty core state) disturbs the Fermi sea
of the conduction electrons so strongly that the subsequent relaxation
into the new ground state follows a fractional power law in time
rather than the usual exponential dependence. This is due to the fact
that the ground states of the initial state and the final state are
orthogonal in the limit of an infinite system (``orthogonality
catastrophe''). At finite, but small $\epsilon$ the relaxation
process involves excitation of a large number of particle-hole pairs
out of the Fermi sea of conduction electrons. A similar situation arises
at the X-ray absorption threshold. There it has been argued that in addition 
to the above an excitonic effect appears, as first discussed by Mahan
\cite{Mahan67}. A theoretical
description requires the use of infinite order perturbation theory.

The problem is in some sense the simplest situation in which strong
electron correlations are generated by a sudden change of electron
occupations of a level coupled to a Fermi sea. The same generic
problem is at the heart of the Kondo problem, or generally speaking,
of quantum impurity problems, which can be understood as a succession
of X-ray edge problems generated by successive flips of the impurity
spin or pseudospin. In an even more general context, such problems
arise in lattice models of correlated electrons, when the hopping of
an electron from one site to the next changes the occupation of these
sites, causing a corresponding rearrangement of the whole Fermi
system. Given the existing evidence that high temperature
superconductors, heavy fermion compounds and other metallic systems
are governed by strong electron correlation effects, which are at
present only poorly understood, there is an urgent need for generally
applicable theoretical methods capable of dealing with these complex
situations.

A powerful method of many-body physics, which directly addresses the
consequences of a change in occupation number of a local level is the
pseudoparticle representation \cite{Barnes76,Cole84}. Within this
framework one introduces pseudoparticles for each of the states of
occupation of a given energy level, i.e.~fermions for the singly
occupied level and bosons for the empty level. It is well known that a
representation of this type for the infinite $U$ Anderson model of a
magnetic impurity in a metal can give surprisingly good results
already in second-order self-consistent perturbation theory
[``non-crossing-approximation'' (NCA)] in the hybridization of local
level and conduction band \cite{NCA}.  However, at low temperatures
and low energies the NCA fails to control the infrared singular
behavior of the pseudoparticle spectral functions at threshold.
Application of the NCA to the problem of the core hole spectral
function gives a threshold exponent $\alpha_d$ independent of the
occupation of the core state, in contradiction with the exactly known
result.

We have recently developed an approximation scheme, which appears to
overcome the difficulties of NCA \cite{Kroha97,Kroha98}. It is based
on the idea of including singular behavior emerging in any of the
two-particle channels. There are two relevant channels, the
pseudofermion-conduction electron and the slave boson-conduction
electron channel. In both channels the ladder diagrams are summed, the
resulting $T\/$-matrices are self-consistently included in the
self-energies, as is required within a conserving approximation
scheme. The main results of this conserving $T\/$-matrix
approximation (CTMA) are: ($i$) the (exactly known) infrared threshold
exponents of the pseudoparticle spectral functions are recovered \cite{Kroha97},
($ii$) the thermodynamic quantities spin susceptibility and specific
heat show local Fermi liquid behavior in the single channel 
case \cite{Kroha05} and
($iii$) in the multi channel case, non-Fermi liquid behavior is 
found \cite{Kroha05},
in quantitative agreement with exact results available in certain
limiting cases.

One of the most stringent tests of a many-body method is the
calculation of the core hole spectral function. In this paper we
report the results of an application of the CTMA to this problem.

The organization of the paper is as follows. In section
\ref{physical_model}, we summarize the most important results of the
exact solution of the X-ray model \cite{Noz3,Schotte69}, notably those
for the threshold exponents for the photoemission and the X-ray
absorption.  Then, in section \ref{representation}, we recall the
pseudoparticle representation of a spinless Anderson impurity
Hamiltonian \cite{Menge88} and point out its equivalence to the X-ray
model in the infrared limit. The conserving pseudopartcle approximation 
up to infinite order in the hybridization $V$ is discussed in section
\ref{theory} and compared with the parquet equation approach of
Nozi{\`e}res {\em et al.\/} \cite{Noz2} in section \ref{comparison}.
The numerical results are discussed in section \ref{results}. In
appendix \ref{ctma_equations} we give explicitly the self-consistent
equations which determine the auxiliary particle self-energies within
the CTMA.

\section{Physical model}
\label{physical_model}

The absorption of an X-ray photon by a deep level core electron and
the subsequent emission of the electron leaves a core hole, which is
seen by the conduction electrons as a suddenly created screened
Coulomb potential. The simplest model Hamiltonian describing this
situation is given by \cite{Mahan67,Noz3,Noz2,Ohtaka90,Noz1}
\begin{equation}
\label{Xray_Hamiltonian}
H=\sum_{\V{k}\sigma} \left( \epsilon^{\phantom{\dag}}_{\V{k}}
    -\mu \right) c^\dag_{\V{k}\sigma} c^{\phantom{\dag}}_{\V{k}\sigma}
    + E^{\phantom{\dag}}_d d^\dag d   + V_d \sum_{\sigma} 
    c^\dag_{0\sigma} c^{\phantom{\dag}}_{0\sigma}  d d^\dag_{\phantom 0} \;,
\end{equation}
where $c^{\phantom{\dag}}_{\V{k}\sigma}$ ($c^\dag_{\V{k}\sigma}$) are
the conduction electron field operators for momentum and spin
eigenstates $|\V{k} \sigma \rangle$, with energy $\epsilon_{\V{k}}$
and chemical potential $\mu$. The energy of the deep level is $E_d$,
and $V_d$ is the screened Coulomb interaction between the conduction
electrons at the site of the hole ($c^\dag_{0\sigma}$,
$c^{\phantom{\dag}}_{0\sigma}$) and the hole (with operators $d^\dag$,
$d$; the spin state of the hole is irrelevant here). We assume that
the hole is localized and does not have internal structure, i.e.~we
neglect the finite life time of the hole due to Auger effect as well
as a possible recoil of the hole.  The Coulomb interaction between the
conduction electrons is absorbed into a quasiparticle renormalization.

\paragraph*{Photoemission.---}
The spectral function of the hole, $A_d(\epsilon)$, which can be
measured in photoemission experiments, is obtained from the
one-particle core hole Green's function $G_d(t)=-i \langle
T[d(t)d^\dag(0)] \rangle$, subjected to the initial condition that the
core hole occupation number $d^\dag d = 0$ for times $t<0$ (before the
photoemission process), by taking the imaginary part of its Fourier
transform, $A_d(\omega)=(1/\pi)\mbox{Im}G_d(\omega-i0)$. The initial
condition is equivalent to the trace $\langle \cdots \rangle$ in the
definiton of $G_d(t)$ being taken only over states with hole
occupation equal to zero. It is this restriction which implies the
non-trivial dynamics of the X-ray problem.  $A_d(\omega)$ is
proportional to the spectral weight of processes, where a photon is
absorbed by the metal, subsequently emitting the deep level core
electron.  The energy $\omega$ required for this process is bounded
from below by the threshold energy $E_0=E_F-E_{\mbox{\scriptsize
    core}}-\Delta E$, where $E_{\mbox{\scriptsize core}}$ and
$E_F$ are the core level energy and the Fermi energy, respectively,
and $\Delta E$ is a renormalization due to core hole-conduction
electron interactions.  In the following we will choose the zero of
energy such that $E_0=0$ (i.e.~$\epsilon=\omega-E_0$). The spectral
function $A_d(\epsilon)$ then shows singular threshold behavior
\begin{equation}
A_d(\epsilon)=\frac{C_d}{\epsilon^{\alpha_d}} \qquad (\epsilon \to 0^+) \; .
\end{equation}
In a landmark paper Nozi{\`e}res and De Dominicis \cite{Noz3}
showed that the exponent $\alpha_d$ depends only on the scattering phase
shift $\eta$ of the conduction electrons off
the core hole and calculated it as ($s\/$-wave-scattering) 
\begin{eqnarray}
\alpha_d = 1-\left( \frac{\eta}{\pi} \right)^2 
         = 1-n^2_d \; ,
\end{eqnarray}
where Friedel's sum rule $\eta=\pi n_d$ has been used to express
$\eta$ in terms of the occupation number of the core level, $n_d$. 

\paragraph*{X--ray absorption.---}
The X-ray absorption cross section is given by the two particle
Green's function $G_2(t)=-i\Theta(t) \langle [ d^\dag_{\phantom 0}(t)
c^{\phantom{\dag}}_{0\sigma}(t), c^\dag_{0\sigma}(0) d(0) ] \rangle$
as $d\sigma/d\epsilon \propto \mbox{Im} G_2(\epsilon-i0)$. The absorption
cross section is finite for $\epsilon > 0$ and again shows singular
threshold behavior
\begin{equation}
\frac{d\sigma}{d\epsilon} = \frac{C_a}{\epsilon^{\alpha_a}} 
                              \qquad (\epsilon \to 0^+) \; .
\end{equation}
The exponent $\alpha_a$ has been calculated by Nozi{\`e}res and De
Dominicis \cite{Noz3} with the result
\begin{eqnarray}
\alpha_a = \frac{2\eta}{\pi} - \left( \frac{\eta}{\pi} \right)^2 
         = 2n_d - n^2_d \; .
\label{absorption_exponent}
\end{eqnarray}

\section{Pseudoparticle representation of the X-ray model}
\label{representation}

As will be seen below, it is useful to formulate the core hole problem in
terms of pseudoparticles in order to impose the initial condition. We define
fermion operators $f^+$ ($f$) and boson operators $b^+$ ($b$) creating
(annihilating) the occupied or empty core level. The transition amplitude $V$
of an electron from the core level into the conduction band describes the
hybridization of these two systems. The Hamiltonian of this system takes the
form of an Anderson impurity Hamiltonian for spinless particles (spin
degeneracy $N=1$):
\begin{eqnarray}
\label{Anderson_Hamiltonian}
H&=&\sum_{\V{k}} \left( \epsilon^{\phantom{\dag}}_{\V{k}} - \mu \right) 
   c^\dag_{\V{k}} c^{\phantom{\dag}}_{\V{k}} \\
 &+& E^{\phantom{\dag}}_d f^\dag f + V \left(
   f^{\dag}_{\phantom{0}}bc^{\phantom{\dag}}_0 
   + \mbox{h.~c.} \right)
   + \lambda Q \; ,         \nonumber
\end{eqnarray}
where $c_0=\sum_{\V{k}}c_{\V{k}}$ annihilates a conduction electron at
the impurity site. The constraint $Q=f^{\dag}f+b^{\dag}b=1$ ensuring
that the core level is either empty or occupied is implemented by
adding the last term in (\ref{Anderson_Hamiltonian}), where $\lambda$
is associated with the operator constraint $Q=1$ and may be
interpreted as the negative of a chemical potential for the
pseudoparticles \cite{Cole84}. As has been shown previously
\cite{Cole84,Kroha98,Costi96}, the limit
$\lambda \to \infty$ imposes the constraint exactly
and is equivalent to taking all expectation values of pseudoparticle
operators in the Hilbert subspace with 
$Q=0$ (no core hole present). Thus, in the present context,
it implements exeactly 
the X-ray initial condition of sudden creation of the core 
hole. The auxiliary particle Green's functions are expressed in terms of 
their self-energies as
$G^{-1}_{f}(i\omega _n) = \left[ G^0_{f,b}(i\omega _n) \right]^{-1} -
\Sigma_{f}(i\omega _n)$,
$G^{-1}_{b}(i\nu _m) = \left[ G^0_{b}(i\nu _m) \right]^{-1} -
\Sigma_{b}(i\nu _m)$, where $G^0_f (i\omega _n) =1/(i\omega _n -E_d)$ and 
$G^0_b (i \nu _m) =1/i\nu _m $ are the
respective non-interacting Green's functions and 
$i\omega _n=(2n+1)\pi/\beta$, $i\nu _m=2m\pi/\beta$ denote
the fermionic and bosonic Matsubara frequencies.

In the model (\ref{Anderson_Hamiltonian}) one may distinguish two
distinct regimes, where the impurity occupation number $n_d$ 
at infinitely long time after suddenly switching on the interaction
is large ($n_d \to 1$, $E_d<0$) or small ($n_d \to 0$, $E_d>0$).
Since, due to the hybridization, $n_d$ is equal and opposite in sign
to the change of the conduction electron number (i.e. screening charge)
induced by the presence of the impurity, $n_d=-\Delta n_c$, 
these regimes correspond via the Friedel sum rule to large
($\eta \to \pi$) and small ($\eta \to 0$) scattering phase shifts,
respectively (see detailed discussion below), and may, therefore, 
be termed the strong and the weak coupling regions.  
We now show the formal equivalence between the X-ray model 
Eq.~(\ref{Xray_Hamiltonian}) and the slave particle Hamiltonian
Eq.~(\ref{Anderson_Hamiltonian}) at low energies 
both in the weak and in the strong coupling regions.

In the strong coupling region, an effective low-energy model
is derived from the Anderson Hamiltonian (\ref{Anderson_Hamiltonian})
by integrating out the slave boson degree of freedom
(or, equivalently, by means of a Schrieffer-Wolff transformation
\cite{SchriefferWolff} onto the part of the Hilbert space
involving only states with the core level occupied).
The interaction term in the resulting effective action reads
\begin{eqnarray}
\label{Seffint}
S_{\mbox{\scriptsize int}}&=&-V^2\frac{1}{\beta^3} 
\sum_{i\omega_n,i\omega'_n,i\nu_m} G_b^0(i\nu_m)\\
&\times& c^\dag_{0}(i\omega'_n-i\nu_m) c^{\phantom{\dag}}_{0}(i\omega_n)
f(i\omega'_n) f^\dag(i\omega_n+i\nu_m)  \; ,     \nonumber
\end{eqnarray}
where, in addition, the projection onto the physical 
Hilbert space
is imposed by taking $\lambda \rightarrow \infty$. 
At low 
\setlength{\unitlength}{1mm}
\begin{figure}
\epsfxsize7.7cm
\centering\leavevmode\epsfbox{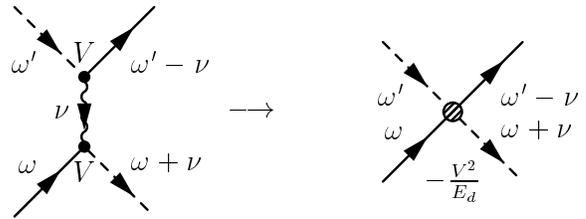}
  \vspace*{0.5cm}
\caption{Diagrammatic representation of the effective 
  low-energy interaction in the strong coupling regime, 
  Eq.~(\ref{Seffint}), and its contraction to a density-density
  interaction at low excitation energies.
  Solid, dashed and wiggly lines correspond here and in the following to 
  conduction electron, pseudo\-fermion and slave boson propagators, 
  respectively.}
\label{fig1}
\end{figure}
\noindent
excitation energy relative to the core level, i.e. when the conduction electron
energies after analytical continuation are 
$|\omega |$, $|\omega'-\nu | \ll |E_d|$ and the 
pseudofermions have energies $\omega'$, $\omega+\nu \approx E_d$
(see Fig.~\ref{fig1}), the non-interacting slave boson Green's
function in Eq.~(\ref{Seffint}) 
is taken at $\nu \approx E_d$ and thus reduces to $1/E_d$.
The resulting effective Hamiltonian is thus 
given by Eq.~(\ref{Xray_Hamiltonian}), with electron operators $d^\dag$, $d$
replaced by pseudofermions $f^\dag$, $f$, interacting with
the conduction electrons via the repulsive, {\it instantaneous}
potential $V_d = -V^2/E_d >0$.

In order to derive the effective low-energy Hamiltonian in the
weak coupling domain ($n_d \to 0$, $E_d>0$), 
it is useful to observe that the model Eq. (\ref{Anderson_Hamiltonian})
is in the physical Hilbert space 
invariant under the special particle-hole transformation
$f\leftrightarrow b$, $c\leftrightarrow c^\dag$ and $E_d \to -E_d$. 
Integrating out the high energy states, i.e. the fermionic degrees of
freedom in this case, and then performing this particle-hole transformation, 
the resulting low-energy Hamiltonian is again given 
by Eq.~(\ref{Xray_Hamiltonian}), with the replacement $d^\dag$, $d$ $\to$
$f^\dag$, $f$, and the attractive interaction potential 
$V_d = -V^2/E_d <0$ between conduction electrons and local pseudofermions.

Having, thus, established the formal connection between the original
X-ray model Eq. (\ref{Xray_Hamiltonian}) and the auxiliary particle
Hamiltonian (\ref{Anderson_Hamiltonian}) 
in the weak and in the strong coupling regions, we now turn
to showing that the photoemission and X-ray absorption spectra are
given by the slave boson and the pseudofermion spectral functions,
respectively.

\paragraph*{Photoemission.---}
The retarded Green's function $G_b^R(t)=-i\Theta(t) \langle [ b(t),
b^\dag(0) ]_- \rangle$ describes the propagation of the empty
$d\/$-level in time. The corresponding spectral function after
projection onto the physical sector $Q=1$, $A_b^+(\omega) = -
\lim_{\lambda \to \infty} \mbox{Im} G_b^R(\omega)/\pi$ can be
represented in terms of the exact eigenstates of the system without
the $d\/$-level, $|0,n\rangle$, and with the $d\/$-level,
$|1,n\rangle$, as \cite{Costi96,Kroha98}
\begin{eqnarray}
A_b(\omega)&=&\\
              \frac{1}{Z_{Q=0}}&\sum_{m,n}&|\langle1,m|b^+|0,n\rangle|^2 
              e^{-\beta \epsilon_{0,n}}\delta(\epsilon+\epsilon_{0,n}
              -\epsilon_{1,m}) \; .   \nonumber
\end{eqnarray}
At zero temperature ($\beta=1/T=\infty$), $A_b(\epsilon)$ is zero for
$\epsilon = \omega -E_0 < 0$, where $E_0=\epsilon_{1,0}-\epsilon_{0,0}$ is the
difference of the ground state energies for the $Q=1$ and $Q=0$
systems. Near the threshold, $\epsilon \gtrsim 0$, $A_b(\epsilon)$ has
a power law singularity (infrared divergence), $A_b(\epsilon) \propto
\epsilon^{-\alpha_b}$, for exactly the same reason as the hole
spectral function $A_d(\epsilon)$ considered above: the states
$|0,n\rangle$ (free Fermi sea) and $|1,n\rangle$ (Fermi sea in
presence of a potential scattering center) are orthogonal, giving rise
to the orthogonality catastrophe \cite{Anderson67}. 
The exponent $\alpha_b$ is therefore given in terms of the phase 
shift $\eta _b$ (for $s\/$-wave scattering) as
$\alpha_b=1-\left( {\eta _b}/{\pi} \right)^2 \;$. Using the Friedel 
sum rule and the fact that in the photoemission process (boson propagator)
the impurity occupation number changes from initially $0$ to $n_d>0$ in the
final state, we obtain the characteristic dependence on $n_d$, 
\begin{equation}
\label{boson_exponent}
\alpha_b=1-n_d^2 \; .
\end{equation}
We may conclude that the threshold behavior of the physical hole
spectral function $A_d(\epsilon)$ and the slave boson spectral
function $A_b(\epsilon)$ is governed by the same exponent,
$\alpha_d=\alpha_b$, provided the scattering phase shift is the same.

\paragraph*{X-ray absorption.---}
In a similar way, the threshold behavior of the X-ray absorption
cross section $d\sigma/d\epsilon$ may be obtained from the pseudofermion 
Green's function. As shown in section \ref{physical_model},
$d\sigma/d\epsilon$ is proportional to the imaginary part of the two
particle Green's function $G_2(t)=-i\Theta(t) \langle [ d^\dag_{\phantom
  0}(t) c^{\phantom{\dag}}_{0\sigma}(t), c^\dag_{0\sigma}(0) d(0) ]
\rangle$. The corresponding quantity here is the slave
boson-conduction electron correlation function
\begin{equation}
G_{bc}(t)=-i\Theta(t) \langle [ b(t)c^{\phantom{\dag}}_0(t), 
          c^\dag_{0\sigma}(0) b^{\dag}_{\phantom{0}}(0)] \rangle \; , 
\end{equation}
which is given in terms of the pseudofermion Green's function
$G_f(\epsilon)$ (after Fourier transformation) as
\begin{equation}
G_{bc}(\epsilon)=\frac{1}{V^2}\left[ \left( G_f^0(\epsilon) \right)^{-1} 
  G_f(\epsilon) - 1 \right] \left( G_f^0(\epsilon) \right)^{-1} \; .
\end{equation}
It follows that the spectral functions are related by
$A_{bc}(\epsilon) \propto A_f(\epsilon) \sim \epsilon^{-\alpha_f}$,
i.e. the X-ray absorption exponent is identical to the pseudofermion threshold
exponent $\alpha _f$.
The latter is again determined by the orthogonality
catastrophe argument, considering that the initial state of the system
is now the conduction electron Fermi sea plus the filled $d\/$-level.
The phase shift $\eta_f$, again given  
via the Friedel sum rule as the change of the occupation number 
from the initial to the final state, is now different,
$\eta_f=(n_d -1)\pi$, leading to the expression
\begin{equation}
\label{fermion_exponent}
\alpha_f=2n_d-n^2_d \; .
\end{equation}
Comparison with (\ref{absorption_exponent}) again shows that the infrared
behavior of the pseudofermion spectral function is indeed identical to that of
the two particle Green's function $G_2$, as expected. 

It should be mentioned that in the intermediate coupling or
``mixed valence'' domain, $\pi N(0) V^2 \approx |E_d|$ 
($n_d \approx 1/2$), a Schrieffer-Wolff type projection 
is no longer valid because of large level occupancy fluctuations. 
The formal derivation of the X-ray
model (\ref{Xray_Hamiltonian}) from the pseudoparticle model
(\ref{Anderson_Hamiltonian}) in the ``mixed valence'' regime
involves a retarded effective interaction, in contrast to 
Eq. (\ref{Xray_Hamiltonian}). However, since 
the  Hamiltonian Eq.~(\ref {Anderson_Hamiltonian}) 
is a faithful representation of a {\it non-interacting} system 
(via the identification $d^{\dag} = f^{\dag}b$), where the 
constraint $Q=f^{\dag}f+b^{\dag}b = 1$ merely serves to implement  
the X-ray initial condition of sudden switching on the interaction
between localized states and the conduction electrons (see above),
the system is described by single-particle wave functions
even in the valence fluctuation regime of this spinless model. 
The analysis of the pseudoparticle threshold exponents 
$\alpha _b$, $\alpha _f$ in terms of the corresponding scattering phase 
shifts $\eta _b$, $\eta _f$ and the Friedel sum rule, as given above, 
then also applies in the valence fluctuation regime. 
It has been verified explicitly by a numerical renormalization group 
calculation of the pseudoparticle threshold exponents that their 
$n_d$ dependence, given in Eqs. (\ref{boson_exponent}), 
(\ref{fermion_exponent}), is valid over the complete 
range of the core level occupation number $n_d$ \cite{Costi94}.

The preceding analysis shows explicitly
that in the auxiliary particle representation
the threshold exponents of both the
X-ray photoemission and absorption are determined by the infrared
behavior of single-particle propagators, involving the physics of the
orthogonality catastrophe for auxiliary bosons or pseudofermions only
\cite{hopfield69,schotte2.69}.
{\it There is no separation into single particle effects and excitonic effects.}

\section{Conserving theory}
\label{theory}
In the previous section we reformulated the core hole problem by 
introducing auxiliary particles and showed on general grounds 
that the threshold exponents of X-ray absorption and 
photoemission spectra can be extracted from one particle properties, namely
the auxiliary fermion and slave boson Green's functions respectively. 
In this section a systematic self-consistent approximation is formulated 
to calculate these functions.

As a minimal requirement the constraint $Q=1$ has to be fulfilled in any
approximate theory. The constraint is closely related to the invariance of the
system under a simultaneous local (in time) gauge transformation $f(\tau) \to
e^{\Theta(\tau)}f(\tau)$, $b(\tau) \to e^{\Theta(\tau)}b(\tau)$. The
Lagrange multiplier $\lambda$ assumes the role of a local gauge field and
transforms as $\lambda \to \lambda + i \partial \Theta / \partial \tau$. Any
approximate scheme respecting the gauge symmetry will preserve the charge $Q$
in time. The simultaneous transformations $f(\tau) \to
e^{\Theta(\tau)}f(\tau)$, $c_{\V{k}}(\tau) \to e^{\Theta(\tau)}c_{\V{k}}(\tau)$,
$\mu(\tau) \to \mu(\tau)+i\partial \Theta / \partial \tau$ lead to the
conservation of the total fermion number $n_f+\sum_{\V{k}} c^\dag_{\V{k}}
c^{\phantom{\dag}}_{\V{k}} = \mbox{const.}$ where $\mu$ is the chemical potential of
the conduction electrons (we choose $\mu=0$). Any theory which
preserves these symmetries is called conserving and may be generated by
functional derivation from a generating functional $\Phi$ of closed skeleton
diagrams \cite{Baym61}.

\paragraph*{NCA. ---}
We are interested in the limit of weak hybridization $V$. So let us
first consider the lowest order approximation. The conserving
approximation scheme requires the self-energies to be determined
self-consistently, 
\begin{figure}
\epsfxsize7.7cm
\centering\leavevmode\epsfbox{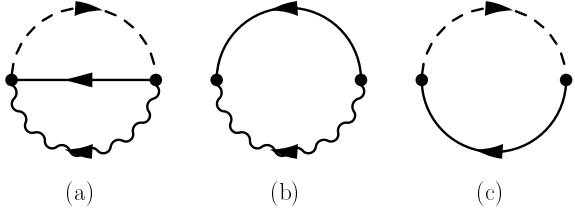}
  \vspace*{0.5cm}
  \caption{(a) Diagrammatic representation of the NCA generating functional.
    (b) and (c) display the pseudo\-fermion and slave boson self-energies
    derived from the NCA functional by functional derivation.}
  \label{fig2}
\end{figure}
\noindent which amounts to an infinite resummation of
perturbation theory even if only the lowest order skeleton diagram ist
kept (which is known as the ``non-crossing-approximation'' (NCA)
\cite{NCA}, see Fig.~\ref{fig2}). The NCA is known to yield good
results in the absence of or sufficiently far away from a Fermi liquid
fixed point \cite{Cox93,Kroha98}.  Hence the NCA is not appropriate in
the X-ray problem. The reason is that no parquet diagrams (see 
Fig.~\ref{fig5}) are included in the lowest order approximation.
By functional derivation of $\Phi$ one obtains for
the slave particle self-energies $\Sigma_f=\delta \Phi / \delta G_f$,
$\Sigma_b=\delta \Phi / \delta G_b$ which are diagramatically given in
Fig.~\ref{fig2} and yield the set of coupled integral equations
\begin{eqnarray}
\label{NCA_Gleichungen}
\Sigma_f(\epsilon)&=&V^2\int_{-\infty}^{\infty} \frac{du}{\pi} \,
                 G_b(\epsilon+u)A_c(-u)f(u) \nonumber \\
\Sigma_b(\epsilon)&=&V^2\int_{-\infty}^{\infty} \frac{du}{\pi} \,
                 G_f(u+\epsilon)A_c(u)f(u) \;
\end{eqnarray}
where $A_c(\epsilon)$ is the non-interacting local conduction electron
spectral density.  At zero temperature $T=0$ the integral equations
can be rewritten as ordinary differential equations (with a constant
density of states for the conduction electrons and for $\epsilon \to
0$) \cite{MH84}
\begin{eqnarray}
\frac{\partial}{\partial \epsilon} \frac{1}{A_f(\epsilon)} 
   &\sim& N(0)V^2A_b(\epsilon) \nonumber \\
\frac{\partial}{\partial \epsilon} \frac{1}{A_b(\epsilon)} 
   &\sim& N(0)V^2A_f(\epsilon) \; .
\end{eqnarray}
The solution displays the well-known infrared singularities
$A_{f,b}(\epsilon) \propto \epsilon^{-\alpha_{f,b}}\quad(\epsilon \to
0)$ where $\alpha_{f,b}=1/2$. These exponents obviously differ from
the exact results discussed before [Eqs.~(\ref{boson_exponent}) and
(\ref{fermion_exponent})].

Hence the NCA is not even in qualitative agreement with the exact
Fermi liquid properties of the model; it shows no dependence of the
exponents on the filling factor $n_d$ of the deep level.This is due to
the lack of vertex corrections which have to be included in infinite
orders of perturbation theory, because it can be shown by
power-counting arguments that there are no corrections to the NCA
exponents in any finite order \cite{Cox93}.

\begin{figure}
\epsfxsize7.7cm
\centering\leavevmode\epsfbox{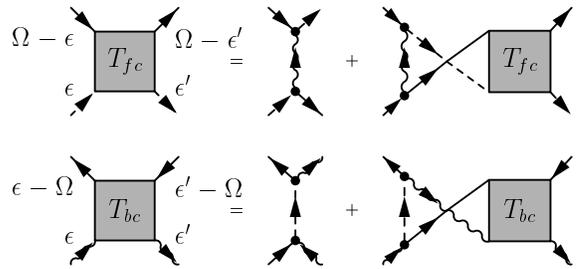}
  \vspace*{0.5cm}
  \caption{Diagramatic representation of the Bethe-Salpeter equations for the
    $T\/$-matrices in Eqs.~(\ref{T_fc-channel}) and (\ref{T_bc-channel}),
    respectively. The analytically continued equations, which are calculated
    numerically, are discussed in appendix \ref{ctma_equations}.}
  \label{fig3}
\end{figure}

\paragraph*{CTMA. ---}
We have to include the major singularities in each order of self-consistent
perturbation theory. These singularities emerge in the conduction electron and
pseudofermion $T\/$-matrix ($T_{fc}$) as well as in the conduction electron
and slave boson $T\/$-matrix ($T_{bc}$).  In order to preserve gauge
invariance, self-consistency has to be imposed: the self-energies are
functionals of the Green's functions which in turn are expressed in terms of
self-energies, closing the set of self-consistent equations.  The summation of
the corresponding ladder diagrams can be performed by solving the integral
equations for the $T\/$-matrices for the pseudofermions (see Fig.~\ref{fig3})
\cite{Kroha97}
\widetext
\top{-2.8cm}
\begin{eqnarray}
T_{fc}(i\omega_n,i\omega'_n;i\Omega_m) &=& V^2 G_b(i\omega_n+i\omega'_n-i\Omega_m)
                                                       \nonumber\\
   &-&\frac{V^2}{\beta}\sum_{i\omega''_n}G_b(i\omega_n+i\omega''_n-i\Omega_m)
    G_f(i\omega''_n)G_c(i\Omega_m-i\omega''_n)
   T_{fc}(i\omega''_n,i\omega'_n;i\Omega_m) \;,
\label{T_fc-channel}
\end{eqnarray}
and the slave-bosons
\begin{eqnarray}
T_{bc}(i\nu_m,i\nu'_m;i\Omega_n) &=& V^2 G_f(i\nu_m+i\nu'_m-i\Omega_n) 
                                                          \nonumber\\
   &-&\frac{V^2}{\beta}\sum_{i\nu''_m}G_f(i\nu_m+i\nu''_m-i\Omega_n) 
    G_b(i\nu''_m)G_c(-i\Omega_n-i\nu''_m)
   T_{bc}(i\nu''_m,i\nu'_m;i\Omega_n) \;.
\label{T_bc-channel}
\end{eqnarray}
\narrowtext
Here $\omega_n, \omega'_n, \omega''_n$ are fermionic frequencies
($\omega_n=(2n+1)\pi/\beta$), $\nu_m, \nu'_m, \nu''_m$ are bosonic
frequencies ($\nu_m=2m\pi/\beta$), and the center of mass frequency
$\Omega_{m,n}$ is bosonic in the case of $T_{fc}$ and fermionic for
$T_{bc}$.  The self-energies $\Sigma_f$ and $\Sigma_b$
\begin{eqnarray}
\Sigma_f(i\omega_n) &=& \Sigma_f^{\mbox{\scriptsize{NCA}}}(i\omega_n) + \Sigma_f^{fc}(i\omega_n)
                        + \Sigma_f^{bc}(i\omega_n) \\
\Sigma_b(i\nu_m) &=& \Sigma_b^{\mbox{\scriptsize{NCA}}}(i\nu_m) + \Sigma_b^{fc}(i\nu_m)
                        + \Sigma_b^{bc}(i\nu_m)
\end{eqnarray}
calculated from $T_{fc}$ and $T_{bc}$, then follow from a generating
functional $\Phi$ (see Fig.~\ref{fig4}) by functional derivation. The
explicit expressions are given in appendix \ref{ctma_equations}.

\begin{figure}
\vspace*{0.4cm}
\epsfxsize7.7cm
\centering\leavevmode\epsfbox{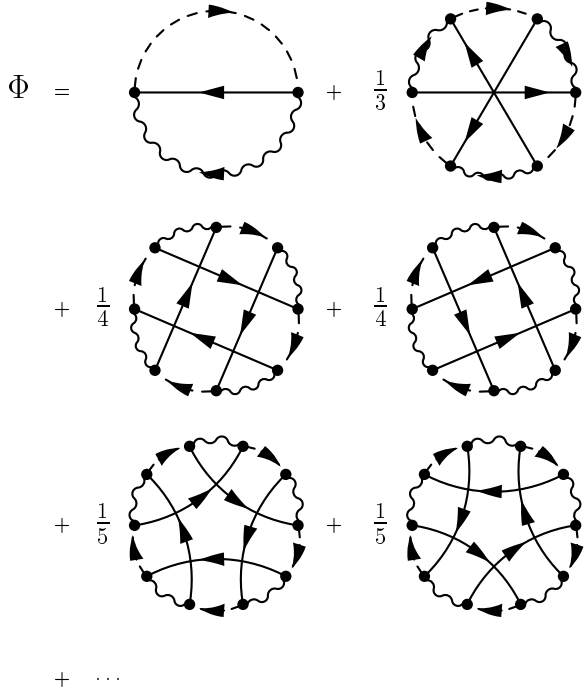}
  \vspace*{0.5cm}
  \caption{Diagrammatic representation of the CTMA generating
    functional. The free energy diagram with two 
    conduction electron lines does not appear, 
    since it is not a skeleton diagram.}
  \label{fig4}
\end{figure}
%

\section{Comparison with renormalized parquet equations}
\label{comparison}

The CTMA is closely related to the parquet equation approach by
Nozi{\`e}res {\em et al.\/} In Ref.~[\cite{Noz1}] these authors investigate the
X-ray model (\ref{Xray_Hamiltonian}) by the methods of perturbation
theory. Even to the lowest order one must sum the so-called parquet
diagrams, in close analogy with the Abrikosov theory of the Kondo
effect \cite{Abrikosov65}.  In this approximation Mahan's prediciton
\cite{Mahan67} of the singularity in the X-ray absorption spectrum was
first confirmed. In a succeeding paper \cite{Noz2} the many-body
approach was generalized to include self-energy and vertex
renormalization in a self-consistent fashion.  This self-consistent
formalism describes the reaction of divergent fluctuations on
themselves, and should, therefore, be useful in other more complicated
problems, such as the Kondo effect.

\begin{figure}
\epsfxsize7.5cm
\centering\leavevmode\epsfbox{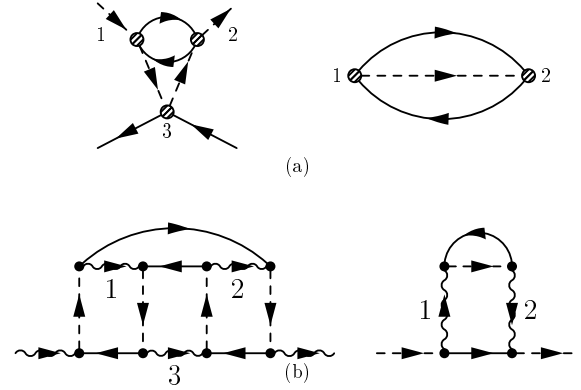}
  \vspace*{0.5cm}
  \caption{(a) Vertex renormalization and self-energy reproduced from 
    the parquet equation approach \protect\cite{Noz2}. These diagrams are 
    obtained from the corresponding ones in (b) by contracting the boson-lines 
    1, 2 and 3. The CTMA, therefore, contains the parquet contributions
    of Ref. [\protect\cite{Noz2}] as a diagrammatic subclass.}
  \label{fig5}
\end{figure}

In Ref.~[\cite{Noz2}] it is shown that the significant contributions in
logarithmic accuracy to the renormalized interaction and the deep level
self-energy are given by the diagrams reproduced in Fig.~\ref{fig5} (a).  
Both graphs are included in the CTMA (see Fig.~\ref{fig5} (b)): 
By collapsing the boson lines into points, i.e. by integrating out      
the high energy bosonic degree of freedom in the strong coupling region
($n_d \to 1$) as done in section \ref{representation}, it is seen that 
the X-ray interaction kernel (Fig. \ref{fig5} (a), left) can be 
extracted from the $T_{bc}\/$-matrix, and
the deep level self-energy (Fig. \ref{fig5} (a), right)
is already included in the NCA.  For weak coupling
($n_d \to 0$) analogous results are obtained by integrating out
the pseudofermionic degree of freedom and then interchanging bosons and 
fermions, compare section \ref{representation}. The {\em self-consistent} 
evaluation of these diagrams represents the renormalized 
parquet analysis for the pseudoparticles. 
{\em The advantage of our formulation is
that it is valid both in the weak coupling and in the strong coupling regime,  
with symmetrical expressions in these two regions}.
The symmetry between weak and strong coupling is also visible in the 
results for the threshold exponents (Fig.~\ref{fig7}).  
Since the CTMA is not
restricted to parquet diagrams (which give the right asymptotic behaviour
only for $V \to 0$), but goes beyond the parquet approximation, one may 
expect that its validity extends beyond the weak and the strong coupling 
limits and interpolates correctly between these regimes.
This will be seen the following section.

\section{Numerical results}
\label{results}

The self-consistent solutions are obtained by first solving the linear
Bethe-Salpeter equations (\ref{T_fc_matrix}) and (\ref{T_bc_matrix})
for the $T\/$-matrices by matrix inversion on a grid of 200 frequency
points. First we insert NCA Green's functions into the $T\/$-matrix
equations. From the $T\/$-matrices the auxiliary particle
self-energies $\Sigma_f$ and $\Sigma_b$ are calculated corresponding
to Eqs.~(\ref{Sigma_f_fc}) and (\ref{Sigma_b_fc}), which give the
respective Green's functions. This process is iterated until
\begin{figure}
\epsfxsize7.7cm
\centering\leavevmode\hspace*{-0.4cm}\epsfbox{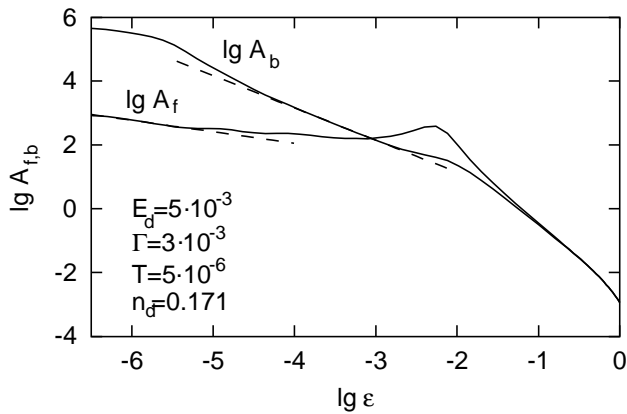}
  \vspace*{0.5cm}
  \caption{Auxiliary particle spectral functions $A_f$ and $A_b$ in the weak 
    coupling regime in a logarithmic plot. 
    The energies are in units of the
    half band width $D$. The slopes of the dashed lines indicate the exact 
    threshold exponents.}
  \label{fig6}
\end{figure}
convergence is reached \cite{convergence}. The $T\/$-matrices show
nonalytic behavior in the infrared limit.

\begin{figure}
\vspace*{0.4cm}
\epsfxsize7.9cm
\centering\leavevmode\hspace*{-0.4cm}\epsfbox{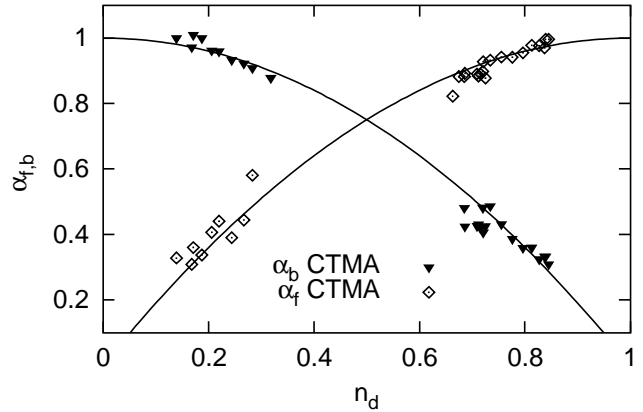}
  \vspace*{0.5cm}
  \caption{Auxiliary particle threshold exponents exctracted from spectra as
    in Fig.~\ref{fig8} for a number of deep level fillings $n_d$. The solid 
    lines represent the exact values derived in Eqs.~(\ref{boson_exponent}) 
    and (\ref{fermion_exponent}).}
  \label{fig7}
\end{figure}

As can be seen from Fig.~\ref{fig6} the fermion and boson spectral
functions display power law behaviour at low frequencies
\cite{small_ed}.  The power law behavior emerges in the infrared
limit, i.e.~for energies smaller than the low energy scale (which is
$E_d$).  For smaller frequencies there is always a deviation from the
power law behaviour due to finite temperature. The exponents
extracted from the spectral functions at low but finite temperature
for various values of the deep
level filling $n_d$ in Fig.~\ref{fig7} are in good numerical agreement with the
exact results in the  regions $n_d \in [0.0,0.3]$ and $n_d \in [0.7,1.0]$. 
Note that in contrast to the $n_d$-dependent exponents within the CTMA the
NCA spectral functions always diverge with $n_d$-independent exponents
$\alpha_f=\alpha_b=1/2$. For intermediate coupling, $n_d \in
[0.3,0.7]$, the convergence of the self-consistent 
\noindent   scheme is very slow, and we find no
stable numerical solution. It remains to be seen whether
this is due to numerical instabilities or possibly due to the
importance of further vertex corrections beyond the CTMA.

\begin{figure}
\vspace*{0.4cm}
\epsfxsize7.9cm
\centering\leavevmode\epsfbox{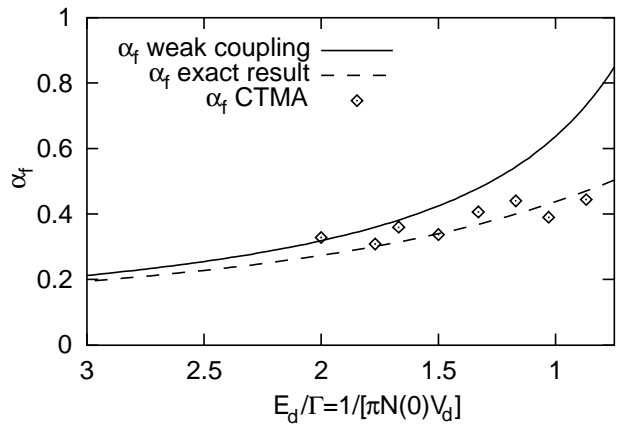}
  \vspace*{0.5cm}
  \caption{Comparison of the CTMA results and the weak-coupling calculation
    \protect \cite{Noz2,Noz1} for the threshold exponent of X-ray absorption spectra.}
  \label{fig8}
\end{figure}

A comparison of the CTMA results with the weak-coupling treatment,
which corresponds to $n_d \to 0$ in our model, shows that for finite
interaction strength renormalization effects are important (see
Fig.~\ref{fig8}). The connection between $n_d$ and $E_d/\Gamma$ is
exactly given by Friedel's sum rule $n_d=1/2-\arctan(E_d/\Gamma)/\pi$.
Again we mention the $n_d$ dependence of the exponent $\alpha_f$ in
contrast to the NCA result: To recover the Fermi liquid properties of
the model one thus has to go far beyond the lowest order
self-consistent approximation.
%

\section{conclusion}

In summary, we have calculated the exponents of threshold singularities
in the X-ray photoemission and absorption spectra, using a standard
many-body technique, where the empty and the singly occupied core level
are represented by separate fields, auxiliary bosons and pseudofermions,
respectively, coupled to the conduction electrons via a hybridization
interaction. In this formulation, the X-ray problem is described by a
spinless Anderson impurity model in pseudoparticle representation, and 
the initial condition of sudden creation of the 
impurity potential is implemented by the constraint
that all expectation values of local fermion or boson fields must be
calculated in the Hilbert subspace with pseudoparticle number
$Q=0$. The latter can be fulfilled exactly. 
It was further shown that the X-ray photoemission cross section or core 
level spectral function is given by the boson spectral function,
while the X-ray absorption cross section is proportional to the total 
fermion hybridization vertex. Therefore, the X-ray photoemission and
absorption threshold exponents are identical to the infrared exponents
of the auxiliary boson and pseudofermion spectral functions, respectively. 
It follows that both X-ray photoemission and 
absorption are solely governed by the orthogonality catastrophe, and
there is no separation into single particle and excitonic effects. 

In a more general context, the generalized SU($N$)$\times$SU($M$) Anderson 
impurity models, classified by the spin degeneracy $N$ of
the local orbital and the number $M$ of degenerate conduction electron channels,
may be considered as standard models to describe strong 
correlations induced by the restriction of no double occupancy of sites. 
Depending on their symmetry, these models display Fermi 
($N=M=1$ or $N\geq M+1$) or non-Fermi liquid behavior ($2\leq N\leq M$) at low 
temperature \cite{Cox93}.
The present case of the spinless Anderson impurity model in slave boson 
representation ($N=1$, $M=1$), Eq. (\ref{Anderson_Hamiltonian}), 
may be considered as the most stringent test case for the development of new 
methods for strongly correlated systems. This is because 
for this case earlier approximation schemes like the 
non-crossing approximation (NCA) fail in the most pronounced way to 
even qualitatively describe the low-energy Fermi liquid behavior of this model,
i.e. the $n_d$ dependence of the infrared threshold exponents, while
in the non-Fermi liquid case the NCA gives the correct exponents at least
in the Kondo limit of these models \cite{Cox93}.

In the present paper we have applied a recently developed approximation scheme, 
the conserving $T\/$-matrix approximation (CTMA) to the 
$N=1$, $M=1$ Anderson impurity model to calculate 
the X-ray photoemission and absorption threshold exponents on a 
common footing. The CTMA includes the complete 
subclass of diagrammatic contributions which, 
in the limits of weak ($n_d \rightarrow 0$) and strong ($n_d \rightarrow 1$) 
impurity scattering potential, reduce to the renormalized parquet diagrams,
which have been shown by Nozi{\`e}res et al.~\cite{Noz2} to 
describe the exact infrared singular behavior in the 
weak coupling regime of the X-ray problem. As a result, the CTMA recovers
the correct X-ray photoemission and absorption exponents in a wide region 
around weak as well as strong coupling. In connection with earlier 
results \cite{Kroha97} on the spin $1/2$ Anderson impurity model 
($N=2$, $M=1$), this makes the CTMA the first standard many-body technique 
to correctly describe the Fermi liquid regime of the Anderson impurity models
in a systematic way, including the smooth crossover to the high temperature 
behavior.

We are grateful for discussions with J.~Brinkmann, T. A. Costi and T.~Kopp.
T.S.~acknowledges the support of the DFG-Graduiertenkolleg ``Kollektive
Ph{\"a}nomene im Festk{\"o}rper''. This work was supported in part by 
SFB 195 of the Deutsche Forschungsgemeinschaft. 
Computer support was provided by the 
John-von-Neumann Institute for Computing, J{\"u}lich.

\appendix
\section{CTMA equations}
\label{ctma_equations}

In this appendix we give explicitly the self-consistent equations
which determine the auxiliary particle self-energies within the CTMA.
In the Matsubara representation the vertex functions $T_{fc}$ and
$T_{bc}$ are given by the following Bethe-Salpeter equations:
\widetext
\top{-2.8cm}
\begin{eqnarray}
\label{Tfc+}
T_{fc}(i\omega_n,i\omega'_n;i\Omega_m) &=& I_{fc}(i\omega_n,i\omega'_n;i\Omega_m)
   \nonumber\\
   &+&\frac{V^2}{\beta}\sum_{i\omega''_n}G_b(i\omega_n+i\omega''_n-i\Omega_m) 
   G_f(i\omega''_n)G_c(i\Omega_m-i\omega''_n)
   T_{fc}(i\omega''_n,i\omega'_n;i\Omega_m) 
\end{eqnarray}
with
\begin{eqnarray*}
I_{fc}(i\omega_n,i\omega'_n;i\Omega_m) = -\frac{V^4}{\beta}\sum_{i\omega''_n}
   G_b(i\omega_n+i\omega''_n-i\Omega_m)G_f(i\omega''_n)
   G_c(i\Omega_m-i\omega''_n)  
   G_b(i\omega'_n+i\omega''_n-i\Omega_m) \; ,
\end{eqnarray*}
and $T_{bc}$
\begin{eqnarray}
\label{Tbc-}
T_{bc}(i\nu_m,i\nu'_m;i\Omega_n) &=& I_{bc}(i\nu_m,i\nu'_m;i\Omega_n)
   \nonumber \\
   &-&\frac{V^2}{\beta}\sum_{i\nu''_m}G_f(i\nu_m+i\nu''_m-i\Omega_n) 
   G_b(i\nu''_m)G_c(i\nu''_m-i\Omega_n)
   T_{bc}(i\nu''_m,i\nu'_m;i\Omega_n) 
\end{eqnarray}
with
\begin{eqnarray*}
I_{bc}(i\nu_m,i\nu'_m;i\Omega_n) = -\frac{V^4}{\beta}\sum_{i\nu''_m}
   G_f(i\nu_m+i\nu''_m-i\Omega_n)G_b(i\nu''_m)
   G_c(i\nu''_m-i\Omega_n) 
   G_f(i\nu'_m+i\nu''_m-i\Omega_n) \; .
\end{eqnarray*}
\bottom{-2.7cm}
\narrowtext
Note that, in addition to the different sign in $T_{fc}$, these
vertex functions differ from the $T\/$-matrices defined before in that
they contain only terms with two or more rungs, since the inhomogenous
parts $I_{fc}$ and $I_{bc}$ represent terms with two bosonic or
fermionic rungs, respectively. The terms with a single rung correspond
to the NCA diagrams and are evaluated separately. 

\begin{figure}
\vspace*{0.4cm}
\epsfxsize7.7cm
\centering\leavevmode\epsfbox{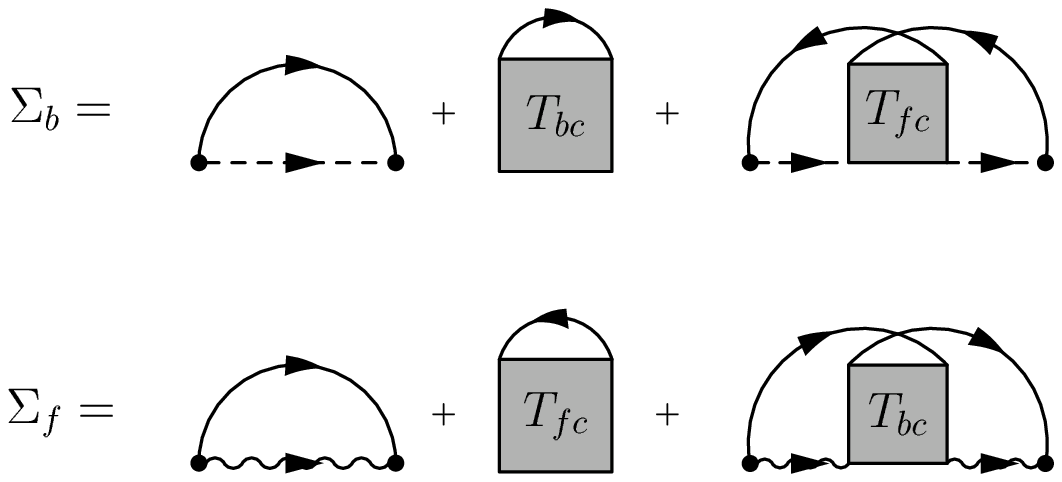}
  \vspace*{0.5cm}
  \caption{Diagrammatic representation of the NCA and CTMA expressions for the
    pseudoparticle self-energies.}
  \label{fig9}
\end{figure}

The fermion self-energies in Fig.~\ref{fig9} are given by
\widetext
\top{-2.8cm}
\begin{eqnarray}
\Sigma_f^{fc}(i\omega_n)&=&\frac{1}{\beta}\sum_{i\Omega_m-i\omega_n}
   G_c(i\Omega_m-i\omega_n)T_{fc}(i\omega_n,i\omega_n;i\Omega_m) 
   \\
\Sigma_f^{bc}(i\omega_n)&=& -\frac{V^2}{\beta^2}\sum_{i\nu'_m,i\nu''_m}
   G_c(i\omega_n-i\nu'_m)G_b(i\nu'_m)G_c(i\omega_n-i\nu''_m)G_b(i\nu''_m)
   T_{bc}(i\nu'_m,i\nu''_m;i\nu'_m+i\nu''_m-i\omega_n) 
\end{eqnarray}
and the boson self-energies by
\begin{eqnarray}
\Sigma_b^{bc}(i\nu_m)&=&\frac{1}{\beta}\sum_{i\nu_m-i\Omega_n}
   G_c(i\nu_m-i\Omega_n)T_{bc}(i\nu_m,i\nu_m;i\Omega_n) 
   \\
\Sigma_b^{fc}(i\omega_n)&=&-\frac{V^2}{\beta^2}\sum_{i\omega'_n,i\omega''_n}
   G_c(i\omega'_n-i\nu'_m)G_f(i\omega'_n)G_c(i\omega''_n-i\nu_m)G_f(i\omega''_n) 
   T_{fc}(i\omega'_n,i\omega''_n;i\omega'_n+i\omega''_n-i\nu_m) \; .
\end{eqnarray}
After analytical continuation to the real frequency axis we have to
solve the NCA equations (\ref{NCA_Gleichungen}) and the following CTMA equations
\begin{eqnarray}
\label{Sigma_f_fc}
\Sigma_f^{fc}(\epsilon) &=& \int_{-\infty}^{\infty} \frac{du}{\pi}\;
   f(u-\epsilon)A_c(u-\epsilon)T_{fc}(\epsilon,\epsilon;u) 
   \\
\label{Sigma_f_bc}
\Sigma_f^{bc}(\epsilon) &=& -V^2 \int_{-\infty}^{\infty} \frac{du}{\pi}
   \int_{-\infty}^{\infty} \frac{du'}{\pi}\; f(u-\epsilon)
   f(u'-\epsilon)A_c(\epsilon-u')G_b(\epsilon) 
   T_{bc}(u,u';u+u'-\epsilon)A_c(\epsilon-u')G_b(u') \\
\label{Sigma_b_bc}
\Sigma_b^{bc}(\epsilon) &=& -\int_{-\infty}^{\infty} \frac{du}{\pi}\;
   f(u-\epsilon)A_c(\epsilon-u)T_{bc}(\epsilon,\epsilon;u) 
   \\
\label{Sigma_b_fc}
\Sigma_b^{fc}(\epsilon) &=& -V^2 \int_{-\infty}^{\infty} \frac{du}{\pi}
   \int_{-\infty}^{\infty} \frac{du'}{\pi}\; f(u-\epsilon)
   f(u'-\epsilon)A_c(\epsilon-u)G_f(\epsilon) 
   T_{fc}(u,u';u+u'-\epsilon)A_c(u'-\epsilon)G_f(u')
\end{eqnarray}
with the fermion-conduction electron vertex function
\begin{eqnarray}
\label{T_fc_matrix}
T_{fc}(\epsilon,\epsilon';\Omega)&=&I_{fc}(\epsilon,\epsilon';\Omega)
-V^2\int_{-\infty}^{\infty}\frac{du}{\pi} \; f(u-\Omega)
   G_b(\epsilon+u-\Omega)G_f(u) A_c(\Omega-u)
   T_{fc}(u,\epsilon';\Omega)  \\
I_{fc}(\epsilon,\epsilon';\Omega)&=&V^4\int_{-\infty}^{\infty}\frac{du}{\pi}\; 
   f(u-\Omega)G_b(\epsilon+u-\Omega)G_f(u)A_c(\Omega-u)
   G_b(\epsilon'+u-\Omega) \nonumber
\end{eqnarray}
and the boson-conduction electron vertex function
\begin{eqnarray}
\label{T_bc_matrix}
T_{bc}(\epsilon,\epsilon';\Omega)&=&I_{bc}(\epsilon,\epsilon';\Omega)
-V^2\int_{-\infty}^{\infty}\frac{du}{\pi} \; f(u-\Omega)
   G_f(\epsilon+u-\Omega)G_b(u) 
   A_c(u-\Omega)T_{bc}(u,\epsilon';\Omega) \\
I_{bc}(\epsilon,\epsilon';\Omega)&=&-V^4\int_{-\infty}^{\infty} \frac{du}{\pi}\;
   f(u-\Omega)G_f(\epsilon+u-\Omega)G_b(x)A_c(u-\Omega)
   G_f(\epsilon'+u-\Omega) \; . \nonumber
\end{eqnarray}
\narrowtext
Note that the self-energy contributions obtained from the two rung 
$T\/$-matrix terms ($I_{fc}$ and $I_{bc}$) display no skeleton diagrams; 
they are subtracted in the end. 

%


\widetext


\begin{references}
\phantom{.}
\vspace*{-10mm}

\bibitem{Anderson67} P.~W.~Anderson, Phys.~Rev.~Lett.~{\bf 18}, 1049 (1967);
  Phys.~Rev.~{\bf 164}, 352 (1967).

\bibitem{Mahan67} G.~D.~Mahan, Phys.~Rev.~{\bf 153}, 882 (1967);
  Phys.~Rev.~{\bf 163}, 612 (1967).

\bibitem{Barnes76} S.~E.~Barnes, J.~Phys.~F. {\bf 6}, 1375 (1976); 
  {\bf 7}, 2637 (1977).

\bibitem{Cole84} P.~Coleman, Phys.~Rev.~B {\bf 29}, 3035 (1984).

\bibitem{NCA} H.~Keiter and J.~C.~Kimball, J.~Appl.~Phys.~{\bf 42}, 1460 (1971);
  N.~Grewe and H.~Keiter, Phys.~Rev.~B {\bf 24}, 4420 (1981);
  Y.~Kuramoto, Z.~Phys.~B {\bf 53}, 37 (1983).

\bibitem{Kroha97} J.~Kroha, P.~W{\"o}lfle and T.~A.~Costi, 
  Phys.~Rev.~Lett.~{\bf 79}, 261 (1997).

\bibitem{Kroha98} J.~Kroha and P.~W{\"o}lfle, Acta Physica Polonica 
  B {\bf 29}, 3781 (1998).

\bibitem{Kroha05} S.~B{\"o}cker, J.~Kroha, and P.~W{\"o}lfle (unpublished).

\bibitem{Noz3} P.~Nozi{\`e}res and C.~T.~De Dominicis, Phys.~Rev.~{\bf 178},
  1097 (1969).

\bibitem{Schotte69} K.~D.~Schotte and U.~Schotte, Phys.~Rev.~{\bf 182},
  479 (1969).

\bibitem{Menge88} B.~Menge and E.~M{\"u}ller-Hartmann, Z.~Phys.~B {\bf 73},
  225 (1988).

\bibitem{Noz2} P.~Nozi{\`e}res, J.~Gavoret, and B.~Roulet, Phys.~Rev.~{\bf 178},
  1084 (1969).

\bibitem{Ohtaka90} For an overview see: K.~Ohtaka and Y.~Tanabe, 
  Rev.~Mod.~Phys.~{\bf 62}, 929 (1990) and references therein.

\bibitem{Noz1} B.~Roulet, J.~Gavoret, and P.~Nozi{\`e}res, Phys.~Rev~{\bf 178},
  1072 (1969).

\bibitem{Costi96} T.~A.~Costi, J.~Kroha, and P.~W{\"o}lfle, 
  Phys.~Rev.~B {\bf 53}, 1850 (1995).

\bibitem{SchriefferWolff} 
  J.~R.~Schrieffer and P.~A.~Wolff, Phys.~Rev.~{\bf 149}, 491 (1966).

\bibitem{Costi94} T. A. Costi, P. Schmitteckert, J. Kroha, and P. W\"olfle,
Physica C {\bf 235-240}, 

\bibitem{hopfield69}
J. J. Hopfield, Comments Solid State Phys.~{\bf 2}, 40 (1969). 

\bibitem{schotte2.69}
K.~D.~Schotte and U.~Schotte, Phys.~Rev.~B {\bf 185}, 509 (1969).

\bibitem{Baym61} G.~Baym, and L.~P.~Kadanoff, Phys.~Rev.~{\bf 124}, 287 (1961); 
  G.~Baym, Phys.~Rev.~{\bf 127}, 1391 (1962).

\bibitem{Cox93} D.~L.~Cox and A.~E.~Ruckenstein, Phys.~Rev.~Lett {\bf 71},
  1613 (1993).

\bibitem{MH84} E.~M{\"u}ller-Hartmann, Z.~Phys.~B {\bf 57}, 281 (1984).

\bibitem{Abrikosov65} A.~A.~Abrikosov, Physics {\bf 2}, 5 (1965).

\bibitem{convergence} The criterion of convergence is defined as follows. If
  the relative change of two successive iterations is smaller than 
  $5.0 \cdot 10^{-3}$ the solution is assumed to be converged.
  
\bibitem{small_ed} In the numerical calculations we choose small
  values for $E_d$ and $\Gamma$ of ${\cal O}(10^{-3})$. The reason is
  that the solution of the self-consistent equations is not stable for
  $E_d$ values of the order of the half bandwith $D$; this may be due
  to influence of the high energy cut-off.

\end{references}
\end{document}